\documentclass[aps,pra,twocolumn,showpacs,superscriptaddress]{revtex4-1}

\usepackage{graphicx}
\usepackage[colorlinks,citecolor=blue,urlcolor=blue,linkcolor=blue]{hyperref}
\usepackage{siunitx}	
\usepackage{amsmath}
\usepackage{amssymb}
\usepackage{amsthm}
\usepackage{braket}
\usepackage{booktabs}
\usepackage{footmisc}
\usepackage{mathrsfs}
\usepackage{cleveref}
\usepackage{dsfont}
\usepackage{subfigure}
\usepackage{multirow}

\newcommand{\unit}{\mathds{1}}

\newcommand{\tr}[1]{\mathrm{Tr}\left(#1\right)}
\newcommand{\trp}[2]{\mathrm{Tr}_\mathrm{#1}\left(#2\right)}
\newcommand{\abs}[1]{\left| #1 \right|} 

\newcommand{\bracket}[2]{\ensuremath{\langle#1 \vphantom{#2}| #2\vphantom{#1}\rangle}}

\begin{document}
\title{Simulation of wave-particle duality in multi-path interferometers on a quantum computer}

\author{Mirko Amico }
\thanks{Mirko Amico is currently employed at Q-CTRL inc.}
\affiliation{The Graduate School and University Center, The
City University of New York, New York, NY 10016, USA}

\author{Christoph Dittel}
\email{christoph.dittel@physik.uni-freiburg.de}
\affiliation{Physikalisches Institut, Albert-Ludwigs-Universit{\"a}t Freiburg, Hermann-Herder-Str. 3, 79104 Freiburg, Germany}
\affiliation{EUCOR Centre for Quantum Science and Quantum Computing, Albert-Ludwigs-Universität Freiburg, Hermann-Herder-Str. 3, 79104 Freiburg, Germany}
\affiliation{Institut f{\"u}r Experimentalphysik, Universit{\"a}t Innsbruck, Technikerstr. 25, 6020 Innsbruck, Austria}

\date{\today}

\begin{abstract}
We present an architecture to investigate wave-particle duality in $N$-path interferometers on a universal quantum computer involving as low as $2\log N$ qubits and develop a measurement scheme which allows the efficient extraction of quantifiers of interference visibility and which-path information. We implement our algorithms for interferometers with up to $N=16$ paths in proof-of-principle experiments on a noisy intermediate-scale quantum (NISQ) device using down to $\mathcal{O}(\log N)$ gates and despite increasing noise consistently observe a complementary behavior between interference visibility and which-path information. Our results are in accordance with our current understanding of wave-particle duality and allow its investigation for interferometers with an exponentially growing number of paths on future quantum devices beyond the NISQ era.
\end{abstract}

\maketitle
\section{Introduction}
The duality of particles and waves played a key role in the development of quantum theory and led Niels Bohr to formulate the  celebrated complementarity principle \cite{Bohr-QM-1935,Bohr-DE-1949}, which states that quantum objects hold both wave- and particle-like features in a mutually exclusive fashion. Its fundamental role manifests in various systems ranging from the interrelation between single- and two-body systems \cite{Jaeger-CO-1993,Jaeger-TI-1995,Kaszlikowski-IT-2003,Peng-QC-2005,Dittel-CB-2020} to complex quantum systems \cite{Carino-WP-2019} and systems of many identical particles \cite{Dittel-WP-2018}. Initially, the basic principle was formulated \cite{Wootters-CD-1979,Greenberger-SW-1988,Mandel-CI-1991,Mandel-CI-1991} and experimentally tested \cite{Rauch-SV-1984,Summhammer-SD-1987,Jacques-DC-2008} for a single particle passing through a two-path interferometer, with which-path information governed by different a-priori probabilities of the individual paths \cite{Wootters-CD-1979,Greenberger-SW-1988,Mandel-CI-1991,Mandel-CI-1991}. Later extensions consider equal a-priori probabilities but which-path detectors placed in each arm of the interferometer \cite{Jaeger-TI-1995,Englert-FV-1996,Dittel-WP-2018}, such that which-path information is due to the distinguishability of the which-path detector states as demonstrated in a multitude of experiments \cite{Duerr-FV-1998,Schwindt-QW-1999,Peng-IC-2003,Peng-QC-2005,Yuan-ED-2018,Gao-ET-2018,Schwaller-EE-2020}. In both cases, wave-particle duality ultimately manifests in a complementary behavior of interference visibility and which-path information. In various theoretical approaches these scenarios were generalized from two- to multi-path interferometers \cite{Duerr-QW-2001,Bimonte-ID-2003,Bimonte-CQ-2003,Jakob-CE-2007,Englert-WP-2008,Siddiqui-TS-2015,Bera-DQ-2015,Bagan-RB-2016,Qureshi-WP-2017}. However, realizing multi-path interference possibly in the presence of which-path detectors constitutes a highly elaborate task, such that only little \cite{Mei-CD-2001} has been undertaken to underpin our current understanding of wave-particle duality for an increasing number of paths. 

In contrast to direct experimental implementations, it becomes more feasible to simulate multi-path interferometers on different experimental platforms \cite{Peng-IC-2003,Peng-QC-2005,Yuan-ED-2018}. Indeed, the simulation of intricate quantum processes constitutes one of the key points for the realization of an universal quantum computer \cite{Feynman-SP-1982}, which has already turned out to be fruitful in various simulations and computational tasks \cite{Monz-RS-2016,Kandala-HE-2017,Zhukov-AS-2018,Amico-ES-2019,Havlicek-SL-2019}. From this perspective, it is, thus, natural to ask how wave-particle duality in multi-path interferometers can efficiently be simulated and tested on a universal quantum computer.

In our present contribution we answer this question. We provide a compact architecture to simulate $N$-path interference in the presence of which-path detectors on a universal quantum computer running on no more than $2 \log N$ qubits, and present a quantum algorithms which allows to efficiently extract quantifiers of the interference visibility and which-path information. In particular, we show how to quantify which-path information from $N$ measurements, and extract the multi-path interference visibility introduced in~\cite{Paul-MQ-2017} from a \emph{single} measurement. We further investigate the $N$-path interference visibility from~\cite{Duerr-QW-2001}, which, in contrast, ideally requires an infinite number of measurements, and find a method to obtain this visibility measure only from a finite number of $2^N$ measurements. We implement the proposed scheme in proof-of-principle experiments on the IBM Q 16 Melbourne NISQ device for $N=2,4,8$, and $16$ paths using quantum circuits with down to $\mathcal{O}(\log N)$ gates, and, despite increasing noise, consistently find a complementary behavior between interference visibility and which-path information.

\section{Multi-path interferometer} 
\begin{figure}[t]
\centering
\includegraphics[width=\linewidth]{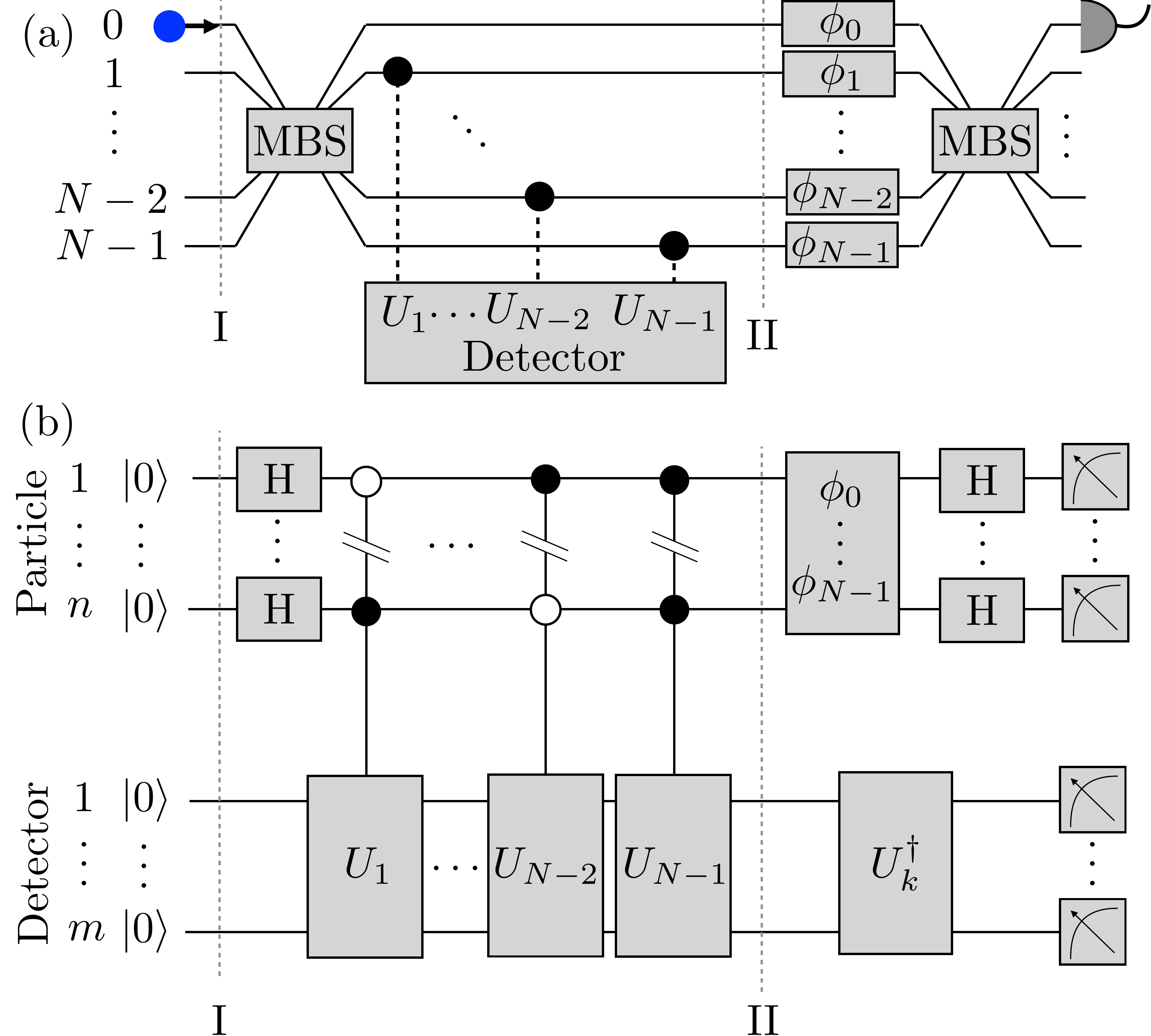}
\caption{Quantum circuit of a $N$-path interferometer. (a) A multiport beam splitter (MBS) sets the particle into a superposition of passing through each path, with a which-path detector acquiring information about the particle's path. Varying phases $\phi_0,\dots,\phi_{N-1}$ give rise to interference fringes after the recombination beam splitter. (b) The particle (resp. which-path detector) in (a) is simulated by the particle (resp. detector) register involving $n=\log N$ (resp. $m\geq n$) qubits. Hadamard gates (H) act as multiport beam splitters, controlled unitary transformations serve to acquire which-path information [here illustrated for the most general case involving $N-1$ operations], with open (resp. filled) circles referring to a control on $\ket{0}$ (resp. $\ket{1}$), and phase gates introduce the phase shifts from (a). In the detector register an additional unitary transformation $U^\dagger_k$ is performed (see main text for details).}
\label{fig:multi-path}
\end{figure}

Let us set the scene as illustrated in Fig.~\ref{fig:multi-path}(a), and consider a single particle entering a $N$-path interferometer in mode $0$, together with a which-path detector that can acquire information about the particle's path. Initially [position $\mathrm{I}$ in Fig.~\ref{fig:multi-path}(a)] we assume an uncorrelated state $\ket{\Psi_\mathrm{I}}=\ket{0}\ket{0}$, with the first (resp. second) ket referring to the particle (resp. which-path detector). The multi-port beam splitter sets the particle into a balanced superposition $1/\sqrt{N} \sum_{j=0}^{N-1} \ket{j}$ of orthogonal states, $\bracket{j}{k}=\delta_{j,k}$, with $\ket{j}$ corresponding to the particle passing through the $j$th interferometer arm. If the particle takes the $j$th arm, the which-path detector can acquire which-path information via the action of the unitary $U_j$ on its initial state. Note that without loss of generality we can set $U_0=\unit$. The action of different unitaries $U_j$ do not necessarily lead to orthogonal which-path detector states, such that the amount of which-path information can be quantified by our ability to discriminate the so obtained states $U_j \ket{0}$ for $j\in\{0,\dots,N-1\}$. Motivated by the upper bound from Ref.~\cite{Feng-UD-2004} on the success probability for unambiguously discriminating these states [see App.~\ref{app:measures}], we consider the quantifier
\begin{align}\label{eq:D}
\mathcal{D}=\sqrt{1- \frac{1}{N(N-1)} \sum_{\substack{j,k=0\\ j\neq k}}^{N-1} \abs{ \bra{0}U_k^\dagger U_j\ket{0}  }^2  }, 
\end{align}
which is built from all mutual overlaps of the which-path detector states, and yields $\mathcal{D}=0$ (resp. $\mathcal{D}=1$) if all states are equal (resp. mutually orthogonal), corresponding to no (resp. full) which-path information.

After position II in Fig.~\ref{fig:multi-path}(a), phase shifters introduce relative phases between the paths, followed by a recombination beam splitter and a measurement of the particle in mode $0$. Due to interference of the particle's paths, varying the phases $\phi=\{\phi_0,\dots,\phi_{N-1}\}$ can lead to a changing probability $p_\mathrm{p}(0|\phi)$ in finding the particle in output $0$. Let us quantify the visibility of this interference effect by maximizing the difference of this probability to its mean $1/N$ over all phase settings,
\begin{align}\label{eq:VC}
\mathcal{V}_\mathrm{C}=&\max_{\phi} \frac{N}{N-1} \left| p_\mathrm{p}(0|\phi) -\frac{1}{N}  \right|.
\end{align}
Interestingly, as shown in detail in App.~\ref{app:measures}, $\mathcal{V}_\mathrm{C}$ relates to the coherence properties of the reduced state of the particle at position II in Fig.~\ref{fig:multi-path}(a) \cite{Paul-MQ-2017}. On the other hand, a similar visibility measure based on the root mean square spread of $p_\mathrm{p}(0|\phi)$ from its mean $1/N$ was introduced in~\cite{Duerr-QW-2001} [Eq.~(1.10) there],
\begin{align}\label{eq:VP}
\mathcal{V}_\mathrm{P} &= \sqrt{ \frac{N^3}{N-1} \left\langle \left( p_\mathrm{p}(0|\phi) - \frac{1}{N}  \right)^2 \right\rangle_\phi },
\end{align}
with $\langle \cdot \rangle_\phi$ the average over all phases $\phi_0,\dots,\phi_{N-1}$. This measure quantifies the purity of the reduced state of the particle at position II in Fig.~\ref{fig:multi-path}(a) [see App.~\ref{app:measures}], and constitutes an upper bound of $\mathcal{V}_\mathrm{C}$ from Eq.~\eqref{eq:VC}, $\mathcal{V}_\mathrm{C} \leq \mathcal{V}_\mathrm{P}$ [see App.~\ref{app:Hierarchy}]. However, both visibility measures are normalized, $0\leq  \mathcal{V}_\mathrm{C},  \mathcal{V}_\mathrm{P} \leq 1$, and together with the which-path information quantifier $\mathcal{D}$ from Eq.~\eqref{eq:D} satisfy the usual wave-particle duality relation
\begin{align}\label{eq:WPD}
\mathcal{D}^2 +\mathcal{V}_\mathrm{C}^2 \leq \mathcal{D}^2 +\mathcal{V}_\mathrm{P}^2=1,
\end{align}
with the inequality saturating if the overlaps $\bra{0}U_k^\dagger U_j \ket{0}$ have equal modulus, and are real or equal for all $j,k\in\{0,\dots,N-1\}$  [see App.~\ref{app:duality}]. Note that while $\mathcal{V}_\mathrm{P}$ and $\mathcal{D}$ are perfectly complementary to each other, for $\mathcal{V}_\mathrm{C}$ the inequality in~\eqref{eq:WPD} does not imply strict complementarity in the sense of an opposite monotonicity behavior.

\section{Quantum circuit} The complementarity between which-path information and interference visibility can be tested on a quantum computer using the circuit from Fig.~\ref{fig:multi-path}(b). The first register comprises $n=\log N$ qubits and is associated with the particle passing through an interferometer with $N$ paths, with $\ket{j}$ corresponding to the particle taking the $j$th path. The which-path detector, on the other hand, is modeled by the second register in Fig.~\ref{fig:multi-path}(b), which contains $m\geq n$ qubits (below we choose $m=n$ in all experiments). In general, the which-path detector may acquire which-path information via \emph{any} unitary operation on the detector register controlled by the state of the particle register. However, in order to account for the most general case, let us consider $N-1$ controlled unitary operations $U_k$ acting on the detector qubits [recall that $U_0=\unit$].

In order to read out the amount of which-path information as quantified by $\mathcal{D}$ from Eq.~\eqref{eq:D}, we additionally act with the unitary $U_k^\dagger$ on the detector register [see Fig.~\ref{fig:multi-path}(b)], and, subsequently, measure this register in the computational basis. This reveals the probability $p_\mathrm{d}(0|k)$ to observe the outcome $\ket{0}$. As we show in App.~\ref{app:Dproof}, by performing this measurement scheme for all unitaries $U_k^\dagger$, with $k\in\{0,\dots,N-1\}$, we obtain $\mathcal{D}$ via
\begin{align}\label{eq:Dm}
\mathcal{D}=\sqrt{ \frac{N}{N-1} \left( 1-\frac{1}{N} \sum_{k=0}^{N-1} p_\mathrm{d}(0|k) \right)  }.
\end{align}

In order to extract the visibilities $\mathcal{V}_\mathrm{C}$ and $\mathcal{V}_\mathrm{P}$ [see Eqs.~\eqref{eq:VC} and~\eqref{eq:VP}] from measurements of the particle register, we introduce phase gates [see Fig.~\ref{fig:multi-path}(b)], such that the particle register state $\ket{j}$ acquires the phase factor $e^{i\phi_j}$, and we denote by $p_\mathrm{p}(0|\phi)$ the probability to observe the particle register in $\ket{0}$ under the phase setting $\phi=\{\phi_0,\dots,\phi_{N-1}\}$. First let us consider $\mathcal{V}_\mathrm{C}$ from Eq.~\eqref{eq:VC}: This visibility measure can be obtained from a \emph{single} phase setting. In particular, for the phases $\phi=\{0,\dots,0\}\equiv 0$ Eq.~\eqref{eq:VC} entails
\begin{align}\label{eq:VCb}
\mathcal{V}_\mathrm{C} \geq \frac{N}{N-1} \abs{p_\mathrm{p}(0|0) -\frac{1}{N}  },
\end{align}
with the inequality saturating in the case of real detector state overlaps $\bra{0}U_k^\dagger U_j\ket{0}$ for all $j,k\in\{0,\dots,N-1\}$ [see App.~\ref{app:measures}]. Thus, in our implementation further below we choose the unitaries $U_j$ such that these overlaps are real, and measure $\mathcal{V}_\mathrm{C}$ via Eq.~\eqref{eq:VCb}.

On the other hand, for $\mathcal{V}_\mathrm{P}$ the phase average in Eq.~\eqref{eq:VP} entails an ideally continuous number of phase settings, which becomes experimentally intractable for an increasing number of paths. However, as we show in App.~\ref{app:VPbproof}, this can be circumvented using the lower bound
\begin{align}\label{eq:VPb}
\mathcal{V}_\mathrm{P} \geq \sqrt{  \frac{N^3}{2^{N+1} (N-1) }  \sum_{\phi\in \{0,\pi\}^N} \left(  p_\mathrm{p}(0|\phi)- \frac{1}{N} \right)^2   },
\end{align}
which also saturates for real overlaps $\bra{0}U_k^\dagger U_j\ket{0}$ of the detector states. Accordingly, $\mathcal{V}_\mathrm{P}$ can be obtained only from $2^N$ different phase settings, which constitutes a vast advantage as compared to the phase average in Eq.~\eqref{eq:VP}.

\section{Simulation results}
We implement our quantum circuit from Fig.~\ref{fig:multi-path} on the IBM Q 16 Melbourne quantum device, and start with $N=2$ paths by utilizing one qubit for each register [see Fig.~\ref{fig:multi-path}(b)]. In this case, the detector register acquires which-path information via the action of a single unitary $U_1$ [recall that $U_0=\unit$], for which we choose a controlled rotation gate $U_1=R_\vartheta$ with angle $\vartheta$, yielding $R_\vartheta\ket{0}=\cos(\vartheta/2) \ket{0} + \sin(\vartheta/2) \ket{1}$ [see App.~\ref{app:circuit} for details]. Thus, in consideration of Eq.~\eqref{eq:D} and the overlap $\bra{0} R_\vartheta\ket{0}=\cos(\vartheta/2)$ between the which-path detector states, the rotation angle $\vartheta$ dictates the amount of which-path information stored in the detector register via $\mathcal{D}=\sqrt{1-\cos^2(\vartheta)}$. On the other hand, for $N=2$ paths we have $\mathcal{V}_\mathrm{C}=\mathcal{V}_\mathrm{P}\equiv \mathcal{V}$. In this case, i.e. for $N=2$, we implement the traditional method of measuring $\mathcal{V}$ [see Eq.~\eqref{eq:VP}], which involves recording interference fringes, and for an increasing number of paths we then utilize our less expensive measurement scheme [see Eqs.~\eqref{eq:VCb} and~\eqref{eq:VPb}] presented above.

\begin{figure}[t]
\centering
\includegraphics[width=\linewidth]{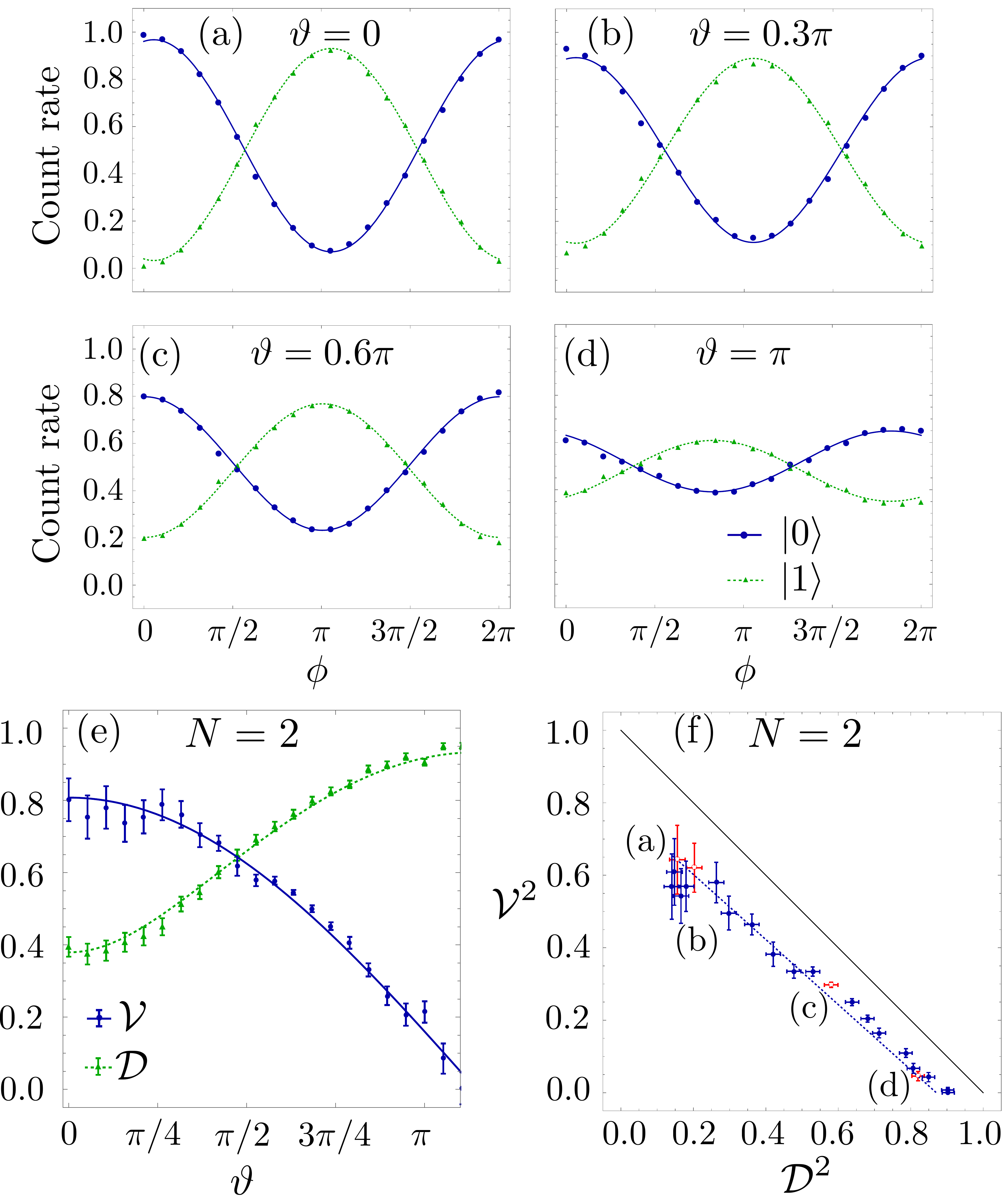}
\caption{Results for a two-path interferometer. For the rotation angles $\vartheta=0$, $0.3\pi$, $0.6\pi$, and~$\pi$, panels (a)-(d) show the interference signal as a function of $\phi$, with the circuit executed $8000$ times for each $\phi$. The normalized count rate of the particle qubit yielding $\ket{0}$ (blue circles) and $\ket{1}$ (green triangles) is fitted by a solid and dotted sine curve, respectively. Error bars arising from counting statistics are too small to be visible. (e) The interference visibility $\mathcal{V}$ (blue circles) -- extracted form the fit to the interference fringes -- and the which-path information quantifier $\mathcal{D}$ (green triangles) is plotted against the rotation angle $\vartheta$, respectively. The solid blue and dotted green line show a fit to the data. (f) For the data in (e) $\mathcal{V}^2$ is plotted against $\mathcal{D}^2$, with the red data points (open circles) corresponding to the interference signals from panels (a)-(d). The dotted blue line results from the fit in (e) and the solid black line illustrates the upper bound from Eq.~\eqref{eq:WPD}. }
\label{fig:2path}
\end{figure}

For varying phases $\phi \in [0,2\pi]$ between the two paths, introduced by a phase gate operating on the qubit of the particle register [see Fig.~\ref{fig:multi-path}(b)], we record interference fringes for $22$ rotation angles $\vartheta$ in the interval $[0,1.1\pi]$ as shown in Fig.~\ref{fig:2path}(a)-(d). For each angle $\vartheta$ we fit the obtained interference signal by a sine curve and extract the interference visibility $\mathcal{V}$ via the fit's amplitude. At the same time, the measurement of the which-path detector qubit reveals the corresponding distinguishability $\mathcal{D}$ via Eq.~\eqref{eq:Dm}. The values of both quantifiers are plotted as a function of the rotation angle $\vartheta$ in Fig.~\ref{fig:2path}(e), and, on the basis of Eq.~\eqref{eq:WPD}, we plot $\mathcal{V}^2$ against $\mathcal{D}^2$ in Fig.~\ref{fig:2path}(f).

Figure~\ref{fig:2path}(e) shows a complementary behavior between interference visibility $\mathcal{V}$ and which-path information $\mathcal{D}$, while the sum of their square lies below unity throughout [see Fig.~\ref{fig:2path}(f)]. This can be explained by noise, imperfect state preparation, and imprecise gate operations, as, for example, evident in the interference signal for $\vartheta=0$ in Fig.~\ref{fig:2path}(a): Noise and imperfect state preparation cause a deviation of $p_\mathrm{p}(0|0)$ from unity, and imprecise Hadamard gates (setting the qubit in an imbalanced superposition) lead to a larger difference of $p_\mathrm{p}(0|\pi)$ from zero as compared to the difference of $p_\mathrm{p}(0|0)$ form unity. Moreover, imperfect rotation gates lead to a non-vanishing interference signal for $\vartheta=\pi$ in Fig.~\ref{fig:2path}(d), with the lowest interference visibility reached for $\vartheta=1.1\pi$. Note that, therefore, we record interference signals for rotation angles $\vartheta\in[0,1.1\pi]$. Moreover, compared to $\vartheta=0$ in Fig.~\ref{fig:2path}(a), we observe a phase shift in the interference signal in Fig.~\ref{fig:2path}(d). This shift, however, does not move the observed signal away from the anti-diagonal line in Fig.~\ref{fig:2path}(f). 

While the IBM Q 16 Melbourne device used for our proof-of-principle experiments has been characterized elsewhere \footnote{According to the specifications \cite{IBM-1,IBM-2} provided by IBM on their website at the time of the experiment, the qubits' average relaxation time $T_1$ and dephasing time $T_2$ is around $54\,\mu$s and $71\,\mu$s, respectively. The average error rates of single-qubit gates and CNOT gates are specified with $0.3\%$ and $7.3\%$, respectively, and the read-out error amounts about $6.6\%$. }, we here model the arising imperfections by accounting for mixed qubits $(1-\epsilon)\ket{0}\bra{0}+\epsilon\ket{1}\bra{1}$ with $\epsilon\in\mathbb{R}$, imbalanced Hadamard gates acting as $\ket{0}\rightarrow T \ket{0}+\sqrt{1-T^2} \ket{1}$ and $\ket{1}\rightarrow \sqrt{1-T^2}  \ket{0}-T\ket{1}$ with $T\in\mathbb{R}$, and a factor $\gamma\in\mathbb{R}$ giving rise to the rotation angle $\gamma\vartheta$ (instead of $\vartheta$). Note that in the ideal case we have $\epsilon=0$, $T=1/\sqrt{2}$, and $\gamma=1$. Fitting our model to the data in Fig.~\ref{fig:2path}(e) reveals the parameters $\epsilon=0.072(3)$, $T=0.767(4)$, and $\gamma=0.873(7)$, and gives rise to the solid lines in Fig.~\ref{fig:2path}(e) and~(f).

We now increase the number of paths to $N=4$, $8$, and $16$, and utilize Eq.~\eqref{eq:VCb} and~\eqref{eq:VPb} to extract the interference visibility $\mathcal{V}_\mathrm{C}$ and $\mathcal{V}_\mathrm{P}$, respectively. For the controlled unitaries $U_j$ [see Fig.~\ref{fig:multi-path}(b)] acting on the detector register, we choose controlled rotation gates, such that a control on the $j$th particle qubit rotates the $j$th detector qubit by the rotation angle $\vartheta$ [see App.~\ref{app:circuit} for details]. Since we choose the same rotation angle $\vartheta$ for all controlled rotation gates, the amount of which-path information obtained by the detector register is fully specified by $\vartheta \in [0,\pi]$, with $\vartheta=0$ (resp. $\vartheta=\pi$) corresponding to no (resp. full) which-path information. This scheme results for all rotation angles $\vartheta$ in real-valued overlaps $\bra{0}U_k^\dagger U_j\ket{0}$ of the resulting which-path detector states, such that the inequalities in Eqs.~\eqref{eq:VCb} and~\eqref{eq:VPb} saturate, and we can utilize them to determine the visibilities $\mathcal{V}_\mathrm{C}$ and $\mathcal{V}_\mathrm{P}$. Note that this scheme involves $\mathcal{O}(\log N)$ rotation gates for the acquisition of which-path information. Hence, for vanishing phase operations (as in the case of investigating $\mathcal{V}_\mathrm{C}$) the total number of gates scales as $\mathcal{O}(\log N)$. In the general case, however, there may be up to $N-1$ controlled unitary operations [cf. Fig.~\ref{fig:multi-path}(b)] requiring at least $\mathcal{O}(N)$ gates.

\begin{figure}[t]
\centering
\includegraphics[width=\linewidth]{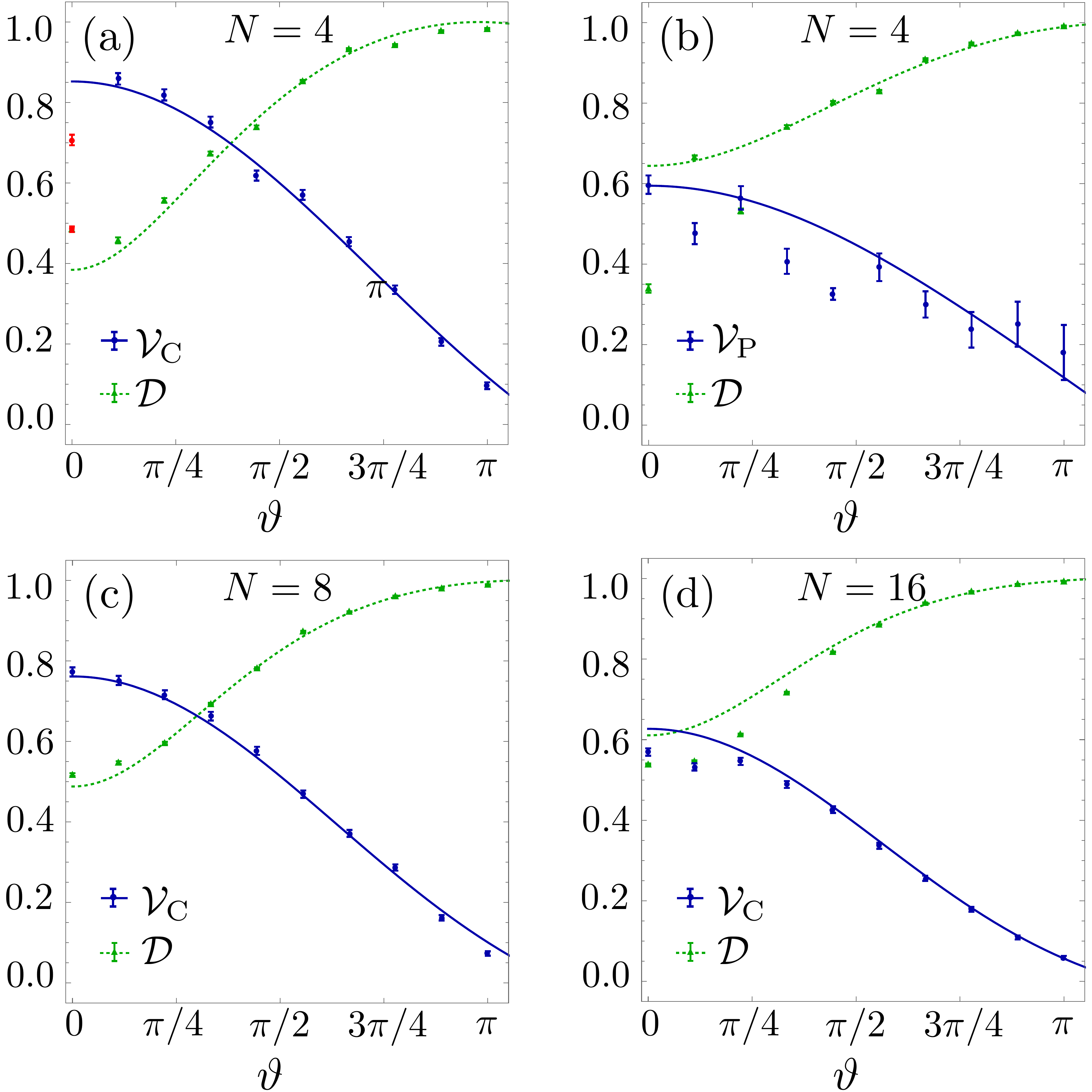}
\caption{Results for interferometers with $N=4$, $8$, and $16$ paths. For $N=4$ paths, panel (a) and (b) show  the interference visibility $\mathcal{V}_\mathrm{C}$ and $\mathcal{V}_\mathrm{P}$ (blue circles) together with the extracted which-path information quantifier $\mathcal{D}$ (green triangles) as a function of the rotation angle $\vartheta$, respectively. Panels (c) and (d) show the recorded date of $\mathcal{V}_\mathrm{C}$ (blue circles) and $\mathcal{D}$ (green triangles) for $N=8$ and $N=16$ paths, respectively. In all panels, solid blue and dotted green lines correspond to a fit to the data, with the red data points for $\vartheta=0$ in (a) excluded due to apparent inconsistencies in the noise level.}
\label{fig:Mpath}
\end{figure}

We implement the circuit for $N=4$ paths using two quibits for both particle and detector register. While $\mathcal{V}_\mathrm{C}$ can be inferred from a single phase setting, the measurement of $\mathcal{V}_\mathrm{P}$ [see Eq.~\eqref{eq:VPb}] requires data collection for $16$ different phases. We perform $8000$ runs for each phase setting, and measure $\mathcal{V}_\mathrm{C}$ and $\mathcal{V}_\mathrm{P}$ together with the which-path information quantifier $\mathcal{D}$ [see Figs.~\ref{fig:Mpath}(a) and~(b)]. By fitting our model accounting for imperfections in the implemented circuit (described for two-paths above) to the data of $\mathcal{V}_\mathrm{C}$, we obtain $\epsilon=0.057(4)$, $T=0.854(13)$, and $\gamma=1.02(3)$. On the other hand, for $\mathcal{V}_\mathrm{P}$ we fix $T=1/\sqrt{2}$ \footnote{Note that we fix $T$ since leaving it as a free parameter would result in a non-vanishing p-value}, and get $\epsilon=0.17(2)$ and $\gamma=0.86(7)$. The difference between the values of $\epsilon$ indicates more noise (due to a larger number of involved gates) in the measurement of $\mathcal{V}_\mathrm{P}$ as compared to $\mathcal{V}_\mathrm{C}$.

As apparent in Figs.~\ref{fig:Mpath}(a) and~(b), extracting $\mathcal{V}_\mathrm{P}$ from $16$ phase settings leads to a considerably larger error as compared to $\mathcal{V}_\mathrm{C}$, which is obtained from a single phase setting only. By further increasing the number of paths to $N=8$ and $N=16$, measuring $\mathcal{V}_\mathrm{P}$ would require data collection for $256$ and $65,536$ different phases, respectively. This, however, leads to insignificant results with the present NISQ device. On the other hand, for the visibility $\mathcal{V}_\mathrm{C}$ we can record data for interferometers as large as $16$ paths, simply limited by the number of available qubits.  The results for $N=8$ paths and $N=16$ paths, using in total $6$ and $8$ qubits, are illustrated in Fig.~\ref{fig:Mpath}(c) and~(d), respectively. Again, we fit our model including imperfections to the data, and obtain $\epsilon=0.075(2)$, $T=0.82(2)$, and $\gamma=0.92(2)$ for $8$ paths, and $\epsilon=0.102(6)$ and $\gamma=0.86(6)$ with a fixed value of $T=1/\sqrt{2}$ \cite{Note2} for $16$ paths.

For a vanishing rotation angle $\vartheta=0$, the noise level characterized by $\epsilon$ gives rise to the difference of the interference visibilities $\mathcal{V}_\mathrm{C}$ and $\mathcal{V}_\mathrm{P}$ from unity, and the difference of $\mathcal{D}$ from zero. Considering this, Figs.~\ref{fig:Mpath}(a), ~(c), and~(d) directly reveal increasing noise for an increasing number of paths, in accordance with $\epsilon$ increasing from $0.057(4)$ over $0.075(2)$ to $0.102(6)$. However, despite imperfections in the circuit, all our results demonstrate an opposite monotonicity behavior [see Fig.~\ref{fig:2path}(e) and Fig.~\ref{fig:Mpath}] and, thus, a complementary behavior between interference visibility and which-path information.

\section{Conclusion} 
Notwithstanding its fundamental role in quantum mechanics, even in the seemingly simple case of a single particle interfering on a multi-path interferometer in the presence of which-path detectors there is only little experimental evidence of wave-particle duality due to the experimentally challenging task of extracting multi-path interference visibilities and which-path information. Here we overcame this challenge and presented a scheme on how to efficiently extract quantifiers of both which-path information and interference visibility. While our scheme allows for an implementation of $N$-path interferometers on diverse experimental platforms operating on as low as $2\log N$ qubits, we here obtained results in favor of our current understanding of wave-particle duality by simulating interferometers with up to $N=16$ paths in proof-of-principle experiments using down to $\mathcal{O}(\log N)$ gates on a NISQ device operating on superconducting transmon qubits. The scaling behavior of our presented quantum circuit (with the number of paths scaling exponentially in the number of qubits) particularly provides an advantage compared to direct implementations of $N$ paths, e.g. via photons in $N$-path optical interferometers. However, while the simulation of wave-particle duality with current available NISQ devices may not provide a computational advantage compared to classical computers, we are optimistic that future quantum devices will go beyond the NISQ era and allow for simulations of wave-particle duality in multi-path interferometers intractable on any classical device.

\begin{acknowledgements}
We acknowledge use of the IBM Q for this work. The authors are grateful to J. Bergou, A. Buchleitner, G. Weihs, and R. Ya. Kezerashvili for valuable and stimulating discussions. C.D. acknowledges the Austrian Academy of Science for a DOC Fellowship, and the Georg H. Endress foundation for financial support. This work began at the 2018 Boulder Summer School for Condensed Matter and Material Physics on Quantum Information. Both authors would like to thank the organizers for the stimulating environment, and the Yale University and the University of Colorado Boulder for hospitality. 
\end{acknowledgements}

\begin{appendix}
\section{Details on $\mathcal{D}$,$\mathcal{V}_\mathrm{C}$, and $\mathcal{V}_\mathrm{P}$}
\label{app:measures}
Under consideration of our scheme from Fig.~\ref{fig:multi-path} we now provide details on the measures $\mathcal{D}$,$\mathcal{V}_\mathrm{C}$, and $\mathcal{V}_\mathrm{P}$. Let us start with considering the common state of the particle and the which-path detector at position $\mathrm{I}$ in Fig.~\ref{fig:multi-path},
\begin{align*}
\rho^\mathrm{I}=\ket{0}\bra{0}\otimes \ket{0}\bra{0},
\end{align*}
with the first (resp. second) term of the tensor product corresponding to the particle (resp. which-path detector). The multi-port beam splitter (or Hadamard gates) then sets the particle into a superposition, $\ket{0} \rightarrow 1/\sqrt{N}\sum_{j=0}^{N-1} \ket{j}$, and the which-path detector acquires information about the particle's path by an action of the unitary $U_j$ if the particle takes the $j$th path. Putting this together, the common state at position $\mathrm{II}$ in Fig.~\ref{fig:multi-path} reads
\begin{align*}
\rho^\mathrm{II}=\frac{1}{N}\sum_{j,k=0}^{N-1}\ket{j}\bra{k} \otimes U_j\ket{0}\bra{0} U_k^\dagger. 
\end{align*}
From this expression we obtain the which-path detector state by tracing over the particle, resulting in
\begin{align}\label{eq:rhod}
\rho_\mathrm{d}= \trp{p}{\rho^\mathrm{II}} =\frac{1}{N} \sum_{j=0}^{N-1} \rho_\mathrm{d,j},
\end{align}
with
\begin{align*}
\rho_\mathrm{d,j} = U_j \ket{0}\bra{0} U_j^\dagger.
\end{align*}
Thus, the which-path detector is in state $\rho_\mathrm{d,j}$ with probability $1/N$, which corresponds to the detection of the particle in the $j$th path. Therefore, the ability to discriminate the states $\rho_\mathrm{d,j}$ (with all states having equal a-priori probability $1/N$) provides a measure for the which-path information obtained by the which-path detector. Let us therefore use the upper bound of the success probability $P_\mathrm{A}$ for unambiguous quantum state discrimination derived in Ref.~\cite{Feng-UD-2004},
\begin{align}
P_\mathrm{A} &\leq 1- \sqrt{ \frac{1}{N(N-1)} \sum_{\substack{j,k=0 \\ j\neq k}}^{N-1} \abs{\bra{0}U_k^\dagger U_j \ket{0} }^2 } \nonumber \\
&\leq \sqrt{ 1- \frac{1}{N(N-1)} \sum_{\substack{j,k=0 \\ j\neq k}}^{N-1} \abs{\bra{0}U_k^\dagger U_j \ket{0} }^2 } \nonumber \\
&\equiv \mathcal{D}, \label{eq:Dapp} 
\end{align}
which motivates the quantifier $\mathcal{D}$ [cf. Eq.~\eqref{eq:D}]. If $\bra{0}U_k^\dagger U_j \ket{0}=1$ for all $j,k\in \{0,\dots,N-1\}$, there is no which-path information, $\mathcal{D}=0$, and, accordingly, a vanishing probability $P_\mathrm{A}$ to discriminate the which-path detector states $\rho_\mathrm{d,j}$. On the other hand, orthogonal which-path detector states, $\bra{0}U_k^\dagger U_j \ket{0}=\delta_{j,k}$, yield a full which-path information, $\mathcal{D}=1$, since the which-path detector states can be discriminated unambiguously. 

Next let us consider the state of the particle at position $\mathrm{II}$ in Fig.~\ref{fig:multi-path}, which we obtain by tracing out the which-path detector, 
\begin{align}\label{eq:rhop}
\rho_\mathrm{p}= \trp{d}{\rho^\mathrm{II}} =\sum_{j,k=0}^{N-1}  [\rho_\mathrm{p}]_{j,k} \ket{j}\bra{k} ,
\end{align}
with
\begin{align}\label{eq:rhopel}
[\rho_\mathrm{p}]_{j,k}=\frac{1}{N} \bra{0}U_k^\dagger U_j \ket{0}.
\end{align}
The phase shifters [see Fig.~\ref{fig:multi-path}] then introduce the phase $e^{i\phi_j}$ in the $j$th path, and the recombination beam splitter (or Hadamard gates) acts on the particle state as $\ket{j} \rightarrow \sum_{j'=0}^{N-1} \beta_{j'} \ket{j'}$, with $|\beta_{j'}|^2=1/N$. Accordingly, the state of the particle after the multi-port beam splitter reads
\begin{align*}
\rho_\mathrm{p,out}=\sum_{j,k=0}^{N-1}  [\rho_\mathrm{p}]_{j,k} e^{i(\phi_j-\phi_k)} \sum_{j',k'=0}^{N-1} \beta_{j'} \beta_{k'}^* \ket{j'}\bra{k'}.
\end{align*}
The probability $p_\mathrm{p}(0|\phi)$ to measure the particle in $\ket{0}$ with the phase setting $\phi=\{\phi_0,\dots,\phi_{N-1}\}$ is then given by
\begin{align}
p_\mathrm{p}(0|\phi)&=\tr{\ket{0}\bra{0} \rho_\mathrm{p,out}  } \nonumber \\
&= \frac{1}{N} \sum_{j,k=0}^{N-1}  [\rho_\mathrm{p}]_{j,k} e^{i(\phi_j-\phi_k)}.\label{eq:pd}
\end{align}

With Eq.~\eqref{eq:pd} at hand, let us now inspect the visibility quantifiers $\mathcal{V}_\mathrm{C}$ and $\mathcal{V}_\mathrm{P}$ provided in Eqs.~\eqref{eq:VC} and~\eqref{eq:VP}. We start with $\mathcal{V}_\mathrm{C}$, which is defined as
\begin{align*}
\mathcal{V}_\mathrm{C}= \max_{\phi} \frac{N}{N-1} \abs{ p_\mathrm{p}(0|\phi)- \frac{1}{N}   }.
\end{align*}
Inserting Eq.~\eqref{eq:pd} yields
\begin{align}
\mathcal{V}_\mathrm{C}&=\max_\phi \frac{1}{N-1} \abs{ \sum_{\substack{j,k=0\\ j\neq k}}^{N-1}  [\rho_\mathrm{p}]_{j,k} e^{i(\phi_j-\phi_k)}}    \nonumber \\
&\leq \frac{1}{N-1} \sum_{\substack{j,k=0\\ j\neq k}}^{N-1} \abs{ [\rho_\mathrm{p}]_{j,k}   }.\label{eq:VCcoh}
\end{align}
Equation~\eqref{eq:VCcoh} highlights that $\mathcal{V}_\mathrm{C}$ is directly related to the (normalized) coherence of the reduced state of the particle $\rho_\mathrm{p}$ [see Eq.~\eqref{eq:rhop}] at position $\mathrm{II}$ in Fig.~\ref{fig:multi-path} \cite{Paul-MQ-2017}. In particular, for the phase setting $\phi=(0,\dots,0)\equiv 0$ we have $e^{i(\phi_j-\phi_k)}=1$, and the inequality in~\eqref{eq:VCcoh} saturates if the overlaps $\bra{0}U_k^\dagger U_j \ket{0}$ of the which-path detector states (and, thus, by Eq.~\eqref{eq:rhopel}, the elements $[\rho_\mathrm{p}]_{j,k}$) are real for all $j,k\in \{0,\dots,N-1\}$.

Next we consider the interference visibility $\mathcal{V}_\mathrm{P}$ from Eq.~\eqref{eq:VP}, defined as
\begin{align*}
\mathcal{V}_\mathrm{P} &= \sqrt{ \frac{N^3}{N-1} \left\langle \left( p_\mathrm{p}(0|\phi) - \frac{1}{N}  \right)^2 \right\rangle_\phi }.
\end{align*}
As shown in Ref.~\cite{Duerr-QW-2001}, by performing the average via a normalized integration over all phases $\phi_0,\dots,\phi_{N-1}$, $\mathcal{V}_\mathrm{P}$ can be expressed in terms of the off-diagonal elements of $\rho_\mathrm{p}$ from Eq.~\eqref{eq:rhop}:
\begin{align}\label{eq:Vp} 
\mathcal{V}_\mathrm{P} = \sqrt{ \frac{N}{N-1}  \sum_{\substack{j,k=0\\j\neq k}}^{N-1} \abs { [\rho_\mathrm{p}]_{j,k} }^2  }.
\end{align}
Utilizing this expression, a short calculation reveals that $\mathcal{V}_\mathrm{P}$ is related to the difference between the purity of $\rho_\mathrm{p}$ and the purity of its incoherent counterpart $\rho_\mathrm{p,inc}=1/N\sum_{j=0}^{N-1}  \ket{j} \bra{j}$, i.e. $\rho_\mathrm{p}$ from Eq.~\eqref{eq:rhop} with zero off-diagonal elements, by
\begin{align*}
\mathcal{V}_\mathrm{P} = \sqrt{ \frac{N}{N-1} \left[  \tr{\rho_\mathrm{p}^2} - \tr{\rho_\mathrm{p,inc}^2} \right]   }.
\end{align*}

\section{Proof of $\mathcal{V}_\mathrm{C}\leq\mathcal{V}_\mathrm{P}$} \label{app:Hierarchy}
In the following, let us prove the hierarchy $\mathcal{V}_\mathrm{C}\leq\mathcal{V}_\mathrm{P}$: Using the Cauchy-Schwarz inequality for the bound of $\mathcal{V}_\mathrm{C}$ from Eq.~\eqref{eq:VCcoh}, we obtain
\begin{align}
\mathcal{V}_\mathrm{C} &\leq \sqrt {\left(  \sum_{\substack{j,k=0\\j\neq k}}^{N-1} \frac{1}{N-1} \abs { [\rho_\mathrm{p}]_{j,k} } \right)^2} \nonumber  \\
&\leq \sqrt {\sum_{\substack{j',k'=0\\j'\neq k'}}^{N-1} \frac{1}{(N-1)^2} \sum_{\substack{j,k=0\\j\neq k}}^{N-1} \abs { [\rho_\mathrm{p}]_{j,k} }^2} \nonumber \\
&=\sqrt {\frac{N}{N-1} \sum_{\substack{j,k=0\\j\neq k}}^{N-1} \abs { [\rho_\mathrm{p}]_{j,k} }^2 } \nonumber \\
&=\mathcal{V}_\mathrm{P} ,\label{eq:hierarchyApp}
\end{align}
where we identified $\mathcal{V}_\mathrm{P}$ from Eq.~\eqref{eq:Vp} in the last step. Note that the first inequality is due to Eq.~\eqref{eq:VCcoh}, and the second inequality saturates if and only if all off-diagonal elements of $\rho_\mathrm{p}$ have equal modulus. 

\section{The duality relation~(\ref{eq:WPD})}
\label{app:duality}
We obtain the duality relation from Eq.~\eqref{eq:WPD} by inserting the elements of $\rho_\mathrm{p}$ from Eq.~\eqref{eq:rhopel} into the expression for $\mathcal{V}_\mathrm{P}$ from Eq.~\eqref{eq:Vp}, yielding
\begin{align}
\mathcal{V}_\mathrm{P}=\sqrt{ \frac{1}{N(N-1)} \sum_{\substack{j,k=0 \\ j\neq k}}^{N-1} \abs{\bra{0}U_k^\dagger U_j \ket{0} }^2 } . \label{eq:Vpoverlap}
\end{align}
In consideration of $\mathcal{D}$ from Eq.~\eqref{eq:D} and the hierarchy~\eqref{eq:hierarchyApp}, we then find
\begin{align*}
\mathcal{D}^2 + \mathcal{V}_\mathrm{C}^2 \leq \mathcal{D}^2 + \mathcal{V}_\mathrm{P}^2 =1,
\end{align*}
which coincides with Eq.~\eqref{eq:WPD}.

\section{Proof of Eq.~(\ref{eq:Dm})} 
\label{app:Dproof}
In the following we prove Eq.~\eqref{eq:Dm}, which provides an expression of the which-path information quantifier $\mathcal{D}$ in terms of the probabilities $p_\mathrm{d}(0|k)$ to measure the detector register in $\ket{0}$ after applying the additional unitary $U_k^\dagger$ [e.g. see Fig.~\ref{fig:multi-path}]: First consider the transformed detector state~\eqref{eq:rhod} at position $\mathrm{II}$ in Fig.~\ref{fig:multi-path}, $U_k^\dagger  \rho_\mathrm{d} U_k$, such that the probability to find the detector register in $\ket{0}$ yields
\begin{align*}
p_\mathrm{d}(0|k)&=\tr{\ket{0}\bra{0} U^\dagger_k \rho_\mathrm{d} U_k}\\
&=\frac{1}{N} \sum_{j=0}^{N-1} \abs{\bra{0} U_k^\dagger U_j \ket{0}}^2.
\end{align*}
Performing this measurement for all $k\in\{0,\dots,N-1\}$, and summing over all probabilities $p_\mathrm{d}(0|k)$ then results in
\begin{align*}
\sum_{k=0}^{N-1} p_\mathrm{d}(0|k) = 1+ \frac{1}{N} \sum_{\substack{j,k=0\\ j\neq k}}^{N-1} \abs{\bra{0} U_k^\dagger U_j \ket{0}}^2.
\end{align*}
Therewith, we obtain $\mathcal{D}$ from Eq.~\eqref{eq:D} through
\begin{align}\label{eq:Dprob}
\mathcal{D}=\sqrt{ \frac{N}{N-1} \left( 1-\frac{1}{N} \sum_{k=0}^{N-1} p_\mathrm{d}(0|k) \right)  },
\end{align}
which proves Eq.~\eqref{eq:Dm}.

\section{Proof of Eq.~(\ref{eq:VPb})} 
\label{app:VPbproof}
In order to prove  Eq.~\eqref{eq:VPb}, reading
\begin{align*}
\mathcal{V}_\mathrm{P} \geq \sqrt{  \frac{N^3}{2^{N+1} (N-1) }  \sum_{\phi\in \{0,\pi\}^N} \left(  p_\mathrm{p}(0|\phi)- \frac{1}{N} \right)^2   },
\end{align*}
we start with its right-hand side abbreviated by $A$, and use $p_\mathrm{p}(0|\phi)=1/N^2 \sum_{j,k=0}^{N-1} \bra{0}U_k^\dagger U_j \ket{0} e^{i(\phi_j-\phi_k)}$ [see Eqs.~\eqref{eq:rhopel} and~\eqref{eq:pd}],
\begin{align*}
A^2&= \frac{N^3}{2^{N+1} (N-1) }  \sum_{\phi\in \{0,\pi\}^N} \left(  p_\mathrm{p}(0|\phi)- \frac{1}{N} \right)^2    \\
=&   \frac{1}{2^{N+1}N (N-1) }  \sum_{\phi\in \{0,\pi\}^N} \left( \sum_{\substack{j,k =0 \\ j\neq k}}^{N-1} \bra{0}U_k^\dagger U_j \ket{0}e^{i(\phi_j-\phi_k)} \right)^2  . 
\end{align*}
By expanding the square of the parenthesis, we obtain
\begin{align}\label{eq:Astep01}
\begin{split}
 A^2=&\frac{1}{2^{N+1}N (N-1) } \sum_{\substack{j,k =0 \\ j\neq k}}^{N-1} \sum_{\substack{j',k' =0 \\ j'\neq k'}}^{N-1} \bra{0}U_k^\dagger U_j \ket{0}  \bra{0}U_{k'}^\dagger U_{j'} \ket{0}  \\
& \times \sum_{\phi\in \{0,\pi\}^N}   e^{i(\phi_j-\phi_k)} e^{i(\phi_{j'}-\phi_{k'})} .
 \end{split}
\end{align}
Here, the third sum yields [note that $j\neq k$ and $j' \neq k'$]
\begin{align*}
\sum_{\phi\in \{0,\pi\}^N}   e^{i(\phi_j-\phi_k)} e^{i(\phi_{j'}-\phi_{k'})} =\begin{cases} 2^N & \text{for } j=j', k=k' \\
2^N & \text{for } j=k', k=j'\\
0 & \text{otherwise.} \end{cases}
\end{align*}
Therefore, Eq.~\eqref{eq:Astep01} becomes
\begin{align}
A^2=& \frac{1}{2N (N-1)}  \sum_{\substack{j,k =0 \\ j\neq k}}^{N-1} \left[ \left( \bra{0}U_k^\dagger U_j \ket{0}\right)^2 + \abs{\bra{0}U_k^\dagger U_j \ket{0}}^2 \right]\nonumber \\
\leq&  \frac{1}{N (N-1)}   \sum_{\substack{j,k =0 \\ j\neq k}}^{N-1} \abs{\bra{0}U_k^\dagger U_j \ket{0}}^2,\label{eq:AboundOvapp}
\end{align}
with the inequality saturating if the overlaps $\bra{0}U_k^\dagger U_j \ket{0}$ of the detector states are real for all $j,k\in\{0,N-1\}$. In consideration of Eq.~\eqref{eq:Vpoverlap}, we then arrive at
\begin{align*}
A^2 \leq \mathcal{V}_\mathrm{P}^2,
\end{align*}
which proves Eq.~\eqref{eq:VPb}. As discussed in the main text, let us stress that for real overlaps of the detector states, Eq.~\eqref{eq:VPb} allows us to measure $\mathcal{V}_\mathrm{P}$ only with $2^N$ phase settings instead of a continuous number of phase settings required for the average in Eq.~\eqref{eq:VP}.

\section{Details on the quantum circuits} 
\label{app:circuit}
\begin{figure}[t]
\centering
\includegraphics[width=\linewidth]{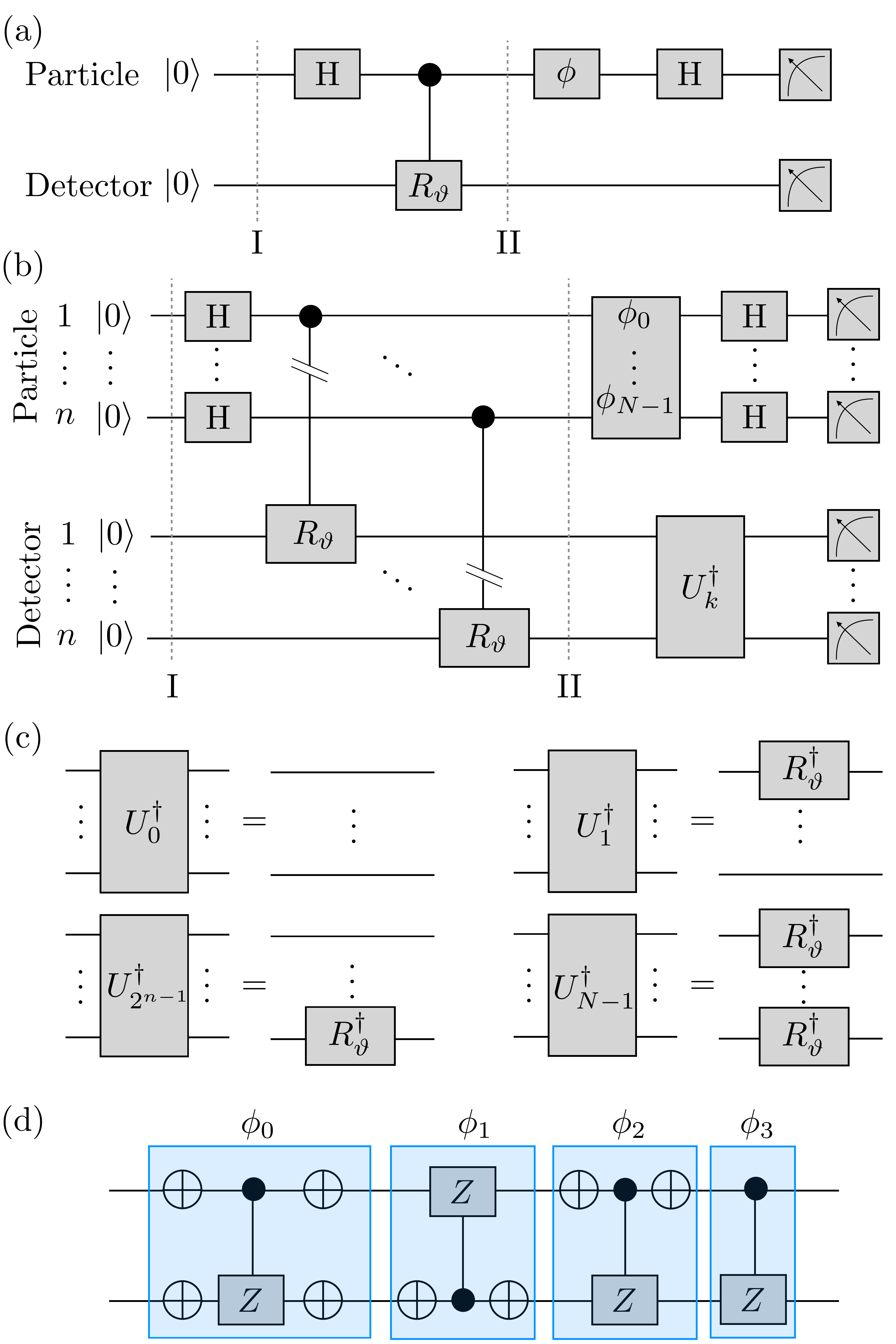}
\caption{Implemented quantum circuits. The data for $N=2$ paths shown in Fig.~\ref{fig:2path} results from the circuit in panel (a). Which-path information is obtained by the detector register via a controlled rotation gate $R_\vartheta$. For $N=4,8$, and $16$ paths, we implement the circuit in panel (b), which gives rise to the data in Fig.~\ref{fig:Mpath} in the main text. Here which-path information obtained by the detector register is also due to controlled rotation gates. The corresponding Hermitian conjugate unitary operations $U_k^\dagger$ are illustrated in panel (c). The phase operations in the particle register in (b) are detailed for $N=4$ path in panel (d). The circuits in the corresponding boxes are activated if $\phi_j=\pi$, and dropped if $\phi_j=0$. }
\label{fig:Circuits}
\end{figure}

The quantum circuit utilized for $N=2$ paths is illustrated in Fig.~\ref{fig:Circuits}(a). It involves a single qubit (i.e. $n=m=1$) for both the particle and detector register. Since we set $U_0=\unit$, there is only a single unitary transformation, $U_1$, through which the detector register (i.e. the detector qubit) can gain information about the particle's path. In particular, for $U_1$ we choose a controlled rotation gate
\begin{align*}
\begin{pmatrix}
1&0&0&0 \\
0&1&0&0 \\
0&0&\cos(\vartheta/2)&-\sin(\vartheta/2) \\
0&0&\sin(\vartheta/2)&\cos(\vartheta/2)
\end{pmatrix},
\end{align*}
which rotates the detector qubit by the angle $\vartheta$ controlled by the particle qubit [see Fig.~\ref{fig:Circuits}(a)]. Since for $N=2$ paths the particle register only comprises a single qubit, the phase operations (setting the relative phases between the states of the particle register in order to record interference fringes) can be performed via a single phase gate on the particle qubit.

In the case of $N=4,8$, and $16$ paths, we also choose controlled rotation gates for the unitaries $U_j$ with which the particle register acquires which-path information. However, for $N=2^n$ paths there are $n$ controlled rotation gates involved, with the $j$th gate rotating the $j$th detector qubit controlled by the $j$th particle qubit [see Fig.~\ref{fig:Circuits}(b)]. Accordingly, the Hermitian conjugate unitaries $U_k^\dagger$ which additionally act on the detector register (in order to read out the amount of which-path information via Eq.~\eqref{eq:Dprob}) correspond to inverse rotations (note that $R_\vartheta^\dagger=R_{-\vartheta}$) as illustrated in Fig.~\ref{fig:Circuits}(c). We note that the quantum circuit and the measurement scheme presented in the main text does not depend on choosing rotation gates for the unitaries $U_j$. However, we choose rotation gates since these gates can easily be implemented on the IBM Q quantum computer and, by their low circuit depth, are subject to little noise. By setting the angle $\vartheta$ equal for all $n$ involved rotation gates, this choice also allows us to investigate the complementarity between the interference visibilities $\mathcal{V}_\mathrm{C}$ and $\mathcal{V}_\mathrm{P}$ and the which-path information quantifier $\mathcal{D}$ as a function of a single parameter, the rotation angle $\vartheta$ [see Figs.~\ref{fig:2path}(e) and~\ref{fig:Mpath}(a)-(d)].

In order to measure the visibility quantifier $\mathcal{V}_\mathrm{P}$ via Eq.~\eqref{eq:VPb}, $2^N$ different phase settings are needed. In particular these phases $\phi=\{\phi_0,\dots,\phi_{N-1} \}$ are given by all combinations $\phi\in\{0,\pi\}^N$. For $N=4$, i.e. for the particle register involving $n=2$ qubits, the quantum circuits to realize these phase operations are illustrated in Fig.~\ref{fig:Circuits}(d): If $\phi_j=\pi$, the circuit in the corresponding box in Fig.~\ref{fig:Circuits}(d) is activated and sets the phase shift, while for $\phi_j=0$ the circuit in the box is dropped. 
\end{appendix}


%

\end{document}